\documentclass[12pt]{article}
\usepackage{epsf,epsfig,psfrag,graphicx}
\usepackage[figuresright]{rotating}
\usepackage{amsmath}
\usepackage{amsfonts}
\usepackage{amssymb}
\usepackage{bbm}
\newcommand{\be}{\begin{equation}}
\newcommand{\ee}{\end{equation}}
\newcommand{\bea}{\begin{eqnarray}}
\newcommand{\eea}{\end{eqnarray}}
\begin{document}
\topmargin 0pt
\oddsidemargin=-0.4truecm
\evensidemargin=-0.4truecm
\renewcommand{\thefootnote}{\fnsymbol{footnote}}
\newpage
\setcounter{page}{0}
\begin{titlepage}
%%%%%%%%%%%%%%%%%%%%%%%%%
%\vspace*{-2.0cm} 
%%%\vspace*{-1.0cm}
%\begin{flushright}
% \\
%%\vspace*{-0.2cm}
%hep-ph/
%\end{flushright}
%%\vspace*{0.5cm}
%\vspace*{0.1cm}
%%%%%%%%%%%%%%%%%%%%%%%%%
%\vspace*{2cm}
\vspace*{1cm}

\begin{center}
{\Large \bf
Do charged leptons oscillate?
%\vspace*{0.16cm} 
%neutrino oscillations
}
\vspace{0.8cm}   

{\large
E. Kh. Akhmedov\footnote{E-mail: akhmedov@mpi-hd.mpg.de}} \\
\vspace{0.25cm}
{\em Max-Planck-Institut f\"ur Kernphysik\\
Postfach 10 39 80, 69029 Heidelberg \\
Germany} \\
\vspace*{2.5mm}
{\em and} \\
\vspace*{2.7mm}
{\em National Research Centre Kurchatov
Institute \\ 123182 Moscow, Russia} \\

\end{center}     
\vglue 0.8truecm

\begin{abstract}
The question of whether charged leptons oscillate is discussed in detail, 
with a special emphasis on the coherence properties of the charged lepton 
states created via weak interactions. This analysis allows one to clarify
also an important issue of the theory of neutrino  oscillations.

\end{abstract}
\vspace{.5cm}   
%%%
%%\centerline{Pacs numbers: 14.60.Pq, 14.60.Lm, 26.65+t}
%\vspace{.5cm}
%\begin{center} Keywords: neutrino oscillations, ... \\
%\end{center}
\end{titlepage}
\renewcommand{\thefootnote}{\arabic{footnote}}
\setcounter{footnote}{0}
\section{Introduction}

Ever since the idea of neutrino oscillations was put forward \cite{Pont,MNS}, 
the question of whether charged leptons can also undergo oscillations has 
been vividly discussed. While most of the authors conclude that such 
oscillations are not possible for one reason or another 
\cite{Pakv,Rich, Dolg,Giu,Burk,Fuji}, others come to the opposite 
conclusion \cite{sriv,Field}. 

Most of the arguments against the oscillations of charged leptons are based 
on the fact that mass eigenstates do not oscillate. This, however, does not 
answer the question of whether certain linear superpositions of charged 
leptons that could in principle be created through weak interactions would 
oscillate into different linear superpositions, leading to observable 
consequences. In the present note we address this question by examining 
the coherence properties of the charged lepton states produced in weak 
interactions. To the best of the present author's knowledge, this issue has 
not been previously studied in the literature. This discussion will also allow 
us to clarify an important issue of the theory of neutrino oscillations, 
namely: Why do neutrinos oscillate? 

\section{Do $e^\pm$, $\mu^\pm$ and $\tau^\pm$ oscillate into each other?}

The answer to this question is the immediate `no', the reason being that 
these charged leptons are mass eigenstates, i.e.  states of definite mass.  
Let us review the simple arguments that show that such particles cannot 
undergo oscillations \cite{Pakv}. 

Assume first that at the time $t_0=0$ and position ${\bf x}_0=0$ a charged 
muon state is created:
\be
|\Psi(0)\rangle=|\mu\rangle\,.
\label{mu}
\ee
After time $t$ and upon propagating the distance $\bf{x}$ this state 
evolves into \footnote{For simplicity, in this section we confine our 
discussion to the plane wave approximation. It is easy to see that a more 
rigorous consideration in terms of wave packets would yield the same result, 
the reason being that the wave packets are normalized.}
\be
|\Psi(t,\,{\bf x})
\rangle~=~e^{-i p_\mu x}|\mu\rangle\,, \qquad\quad
p_\mu x~=~E_\mu t - {\bf p}_{\mu} {\bf x}\,,
\label{muevolv}
\ee
where $E_\mu$ and ${\bf p}_{\mu}$ are the energy and 3-momentum of the muon, 
and for simplicity we have ignored the fact that muon is unstable (this 
is essentially irrelevant to the question we want to address). The 
probability for the muon to remain itself and not to oscillate into 
electron or tauon is then  
\be
%\Rightarrow\quad
P_{\mu\mu}~=~|\langle\mu|\Psi(t,\,{\bf x})\rangle|^2~=~1\,.
\label{Pmu1}
\ee

Consider now the situation where the initially produced charged lepton 
state is a linear superposition of e.g. muon and electron:

\be
|\Psi(0)\rangle~=~\cos\theta|\mu\rangle~+~e^{i\alpha}\sin\theta|e\rangle
\label{mix1}
\ee
with real $\theta$ and $\alpha$. The weights of ~$\mu$ ~and ~$e$ ~in this  
state are $\cos^2\theta$ and $\sin^2\theta$ respectively. The evolved state 
is then
\be
|\Psi(t,\,{\bf x})\rangle~=~e^{-i p_\mu x}\cos\theta |\mu\rangle ~+~
e^{-i p_e x}e^{i\alpha} \sin\theta |e\rangle\,. 
\label{mix1evolv}
\ee
The probabilities of finding $\mu$ and $e$ in the evolved state are
\be
P_{\mu}~=~|\langle\mu|\Psi(t,\,{\bf x})\rangle|^2~=~|e^{-i p_\mu x}
\cos\theta|^2~=~\cos^2\theta\,, \quad
\ee
\be
P_{e}~=~|\langle e|\Psi(t,\,{\bf x})\rangle|^2~=~|e^{-i p_e x+i \alpha} 
\sin\theta|^2~=~\sin^2\theta\,,
\label{PmuPe}
\ee
i.e. the same as in the initial state (\ref{mix1}). Thus, there are no 
oscillations between mass-eigenstate charged leptons $e$, $\mu$ and
$\tau$, no matter if the initial state is pure or a coherently mixed one.
The reason for this is that mass eigenstates evolve by simply picking up 
phase factors whose moduli are always equal to unity.

It should be noted that the same argument applies to neutrinos: an initially 
produced flavor state, say $\nu_e$, can oscillate with some probability into 
$\nu_\mu$ or $\nu_\tau$, but the weights of the mass eigenstates $\nu_1$, 
$\nu_2$ and $\nu_3$ in such a state will not change with time.  

\section{Oscillation between superpositions of $e$, $\mu$ and $\tau$}

A natural question then is: Can we imagine a situation when one 
creates a coherent superposition of $e$, $ \mu$ and $\tau$ and then
also \underline{detects} their coherent superposition (the same or
different) rather than individual mass-eigenstate charged leptons?
%%%%
If this were possible, one would be able to observe oscillations between 
such mixed charged lepton states \cite{Pakv}. 
%%%%

Closely related to the above question is the following one: 
Why do we say that in charged-current weak interactions charged leptons 
are emitted and detected as mass eigenstates and neutrinos as flavor 
states (superpositions of mass eigenstates) and not vice versa? Or not both 
as some superpositions of mass eigenstates?  
After all, charged-current weak interactions are completely symmetric 
with respect to neutrinos and charged leptons, 

\be
{\cal L}_{\rm CC}=-\frac{g}{\sqrt{2}} \left(\bar{e}_{aL} \gamma^\mu
U_{ai}\nu_{iL}\right) W^-_\mu ~+~h.c.\,, \qquad
(a=e, \mu, \tau,~~~i=1, 2, 3)\,, 
\label{Lcc}
\ee

\noindent
with the leptonic mixing matrix $U$ coming from  the diagonalization of the 
mass matrices of both charged leptons and neutrinos, so why cannot charged 
leptons be created and absorbed in weak interactions as coherent 
superpositions of mass eigenstates? What is the origin of the disparity 
between neutrinos and charged leptons?

One might suspect that this disparity comes about because of the enormous
difference between the masses of charged leptons and neutrinos, and as we
shall see, this is indeed the case. However, it is important to understand
how exactly this mass difference comes into play.

Let us consider the problem in more detail. The question we want to address 
is how do we know that a charged lepton emitted or absorbed in a weak 
interaction process is either $e$ or $\mu$ or $\tau$ but not their coherent 
superposition. This actually amounts to asking why neutrinos oscillate, 
because it is the fact that charged leptons participate in weak interactions 
as mass eigenstates that ``measures'' the neutrino flavor, i.e. ensures that 
neutrinos are emitted and captured as well-defined coherent superpositions of 
mass eigenstates 
\footnote{Note that for charged leptons their flavor is {\sf defined} to 
coincide with their mass.}.

In the case of nuclear $\beta$ decay the situation is simple: only 
$e^\pm$ can be emitted together with a neutrino or antineutrino, because 
there is no energy available to produce $\mu^\pm$ or $\tau^\pm$. The same is 
also true for muon decays $\mu^\pm\to e^\pm \nu \bar{\nu}$. Thus, in these 
cases the emitted charged lepton is obviously a pure mass eigenstate.

Consider, however, decays of charged pions $\pi^\pm\to l^\pm \nu$ (or 
similarly for charged kaons). Here the decay energy is sufficient for the 
production of both electrons and muons, i.e. $l=e\,,\mu$. So how do we 
know that the produced charged lepton is either $e$ or $\mu$ and not their 
coherent superposition? As was already pointed out, this is actually the 
same as asking how do we know that the emitted neutrino is either $\nu_e$ or 
$\nu_\mu$. Of course, if e.g. a $\mu^+$ produced in the pion decay is 
detected, than we know that the neutrino born in the same process is 
$\nu_\mu$. But what if the charged lepton is not detected, as it is usually 
the case?  

To illustrate the arising problem, consider a hypothetical situation when 
neutrinos produced or absorbed in weak interactions are mass eigenstates 
$\nu_1$, $\nu_2$ and $\nu_3$, whereas the associated charged leptons are 
\vspace*{-2.0mm}
\bea
|e_1\rangle~=~U_{1e}|e\rangle~+~U_{1\mu}|\mu\rangle~+~U_{1\tau}|\tau\rangle\,,
\nonumber \\
|e_2\rangle~=~U_{2e}|e\rangle~+~U_{2\mu}|\mu\rangle~+~U_{2\tau}|\tau\rangle\,,
\nonumber \\
|e_3\rangle~=~U_{3e}|e\rangle~+~U_{3\mu}|\mu\rangle~+~U_{3\tau}|\tau\rangle
\,,\,
\label{ei}
\eea
which are emitted or detected together with $\nu_1$, $\nu_2$ and $\nu_3$ 
respectively. This possibility is perfectly consistent with the 
charged-current interaction Lagrangian (\ref{Lcc}). However, if this were the 
case, then charged leptons $e_1$, $e_2$ and $e_3$ would oscillate into each 
other, while neutrinos would not be able to oscillate. We know that in reality 
neutrinos do oscillate, so what is wrong with this apparently 
consistent possibility?

To make the problem look even worse, one could conceive a situation in which 
both charged leptons and neutrinos participating in charged-current 
weak interactions are coherent superpositions of their respective mass 
eigenstates:

\be
|e_\beta\rangle = \sum_a W_{\beta a}^*|e_a\rangle\,,\qquad
|\nu_\beta\rangle = \sum_i V_{\beta i}^*|\nu_i\rangle\,, \quad e_a=e, \mu, 
\tau,~~~i=1, 2, 3, 
\label{enui}
\ee
where $W$ and $V$ are $3\times 3$ unitary matrices satisfying the condition 
\be
W^\dag V = U 
\label{cond1}
\ee
but otherwise arbitrary. Eq.~(\ref{enui}) defines the new quantum number 
of neutrinos and charged leptons which we shall call ``odor'' to distinguish 
it from the usual leptonic flavor.  The special case $W=\mathbbm{1}$, $V=U$ 
corresponds to the standard situation where the charged leptons 
participating in weak interactions are mass eigenstates, while neutrinos 
are the flavor eigenstates $\nu_e$, $\nu_\mu$ and $\nu_\tau$, whereas 
the special case $W=U^\dag$, $V=\mathbbm{1}$ corresponds to the situation 
where the weak-eigenstate charged leptons are given by eq.~(\ref{enui}), 
and neutrinos are emitted and absorbed as mass eigenstates.

Had weak interactions selected the neutrino states $\nu_\beta$ defined in 
eq.~(\ref{enui}) as weak eigenstates, then by detecting such a neutrino we 
would measure the odor of the associated charged lepton; in this case 
the charged leptons states $e_\beta$ could oscillate into each other. However, 
these oscillations would only occur if {\sf both} neutrino and charged lepton 
produced in the same decay were detected, i.e. they would be a manifestation of 
an EPR-like correlation~\cite{EPR}. Likewise, for neutrinos to oscillate, one 
would have to measure their odor by detecting the charged lepton state 
emitted in the same decay. At the same time, neutrinos are known to oscillate 
even when the associated charged leptons are not detected. To understand why 
this happens and why charged leptons do not oscillate we have to study the 
coherence properties of the charged lepton states produced in weak interaction
processes.

\section{Coherence properties of charged lepton states}

Unlike neutrinos which can be produced or detected only via weak 
interactions
\footnote{Ignoring possible new interactions responsible for the neutrino 
mass generation.}, 
charged leptons participate also in electromagnetic interactions and are 
usually detected through them. The electromagnetic interactions are, 
however, flavor-blind, and therefore of no interest to us here. 
We shall thus concentrate on the coherence properties of charged lepton 
states produced or detected in weak-interaction processes.

The energy $E$ and momentum $p$ of a particle produced, e.g., in some decay 
process have quantum-mechanical uncertainties, $\sigma_E$ and $\sigma_p$. 
This, in particular, means that the particle should be  described by a wave 
packet of the spatial size $\sigma_x\sim 1/\sigma_p$ rather than by a plane 
wave. The knowledge of the particle's energy and momentum and their 
corresponding uncertainties would allow one to determine the squared mass 
of the particle with an uncertainty $\sigma_{m^2}$.

Let us consider for definiteness $\pi^\pm\to l^\pm \nu$ decays. 
The energies and momenta of the produced charged leptons and their 
corresponding quantum-mechanical uncertainties are determined by the decay 
conditions. If the uncertainty in the inferred mass of the charged lepton 
$\sigma_{m^2}$ satisfies
\footnote{Our arguments here are similar to those used in the discussion of 
the coherence of neutrinos in \cite{Kayser}.}
\be
\sigma_{m^2}< m_\mu^2-m_e^2\,, 
\label{cond2}
\ee
then for for each decay event one would exactly know which particular 
charged lepton was produced. The fact that in each decay 
either $\mu$ or $e$ is produced (with their respective probabilities) 
would imply that the produced charged lepton state is an incoherent 
mixture of $\mu$ and $e$.  If, on the contrary,  
\be
\sigma_{m^2}> m_\mu^2-m_e^2\,, 
\label{cond3}
\ee

\noindent
then it will be in principle impossible to determine which mass-eigenstate 
charged lepton was produced in the decay process. The amplitudes of the 
emission of $\mu$ and $e$ would then add coherently, i.e. the emitted 
charged lepton state would be a coherent superposition of $\mu$ and $e$. 
The situation here is quite similar to that with the electron interference 
in double slit experiments:  If there is no way to find out which slit the 
detected electron has passed through, the detection probability will exhibit 
an interference pattern, but if such a determination is possible, the 
interference pattern will be washed out. 

Now let us estimate the mass uncertainties of charged leptons produced in 
$\pi^\pm \to l^\pm \nu$ decays. 
Assuming that the uncertainties $\sigma_E$ and $\sigma_p$ are uncorrelated, 
from the relativistic relation between the mass, energy and momentum of a 
free particle $m^2=E^2-p^2$ where $p\equiv |\bf{p}|$, one finds 

\vspace*{-2mm}
\be
\sigma_{m^2}=\left[(2 E \sigma_E)^2+(2 p\sigma_p)^2\right]^{1/2}\,.
\label{sigmam1}
\ee

\vspace*{2.5mm}
\noindent
For an isolated decaying particle (or when its interaction with the environment 
can be neglected) the quantum-mechanical uncertainties of the energies of the 
decay products are essentially given by their parent's decay width. Thus, for 
charged leptons produced in $\pi^\pm \to l^\pm \nu$ decays we have 

\vspace*{-3mm}
\be
\sigma_E\simeq \Gamma_\pi=\Gamma_\pi^0/\gamma\,,
\label{sigmae}
\ee

\vspace*{2.5mm}
\noindent
where $\gamma=(1-v^2)^{-1/2}$ is the pion's Lorentz factor and 

\be
\Gamma_\pi^0=2.5\cdot 10^{-8}~{\rm eV}\,
\label{Gammapi0}
\ee

\vspace*{2.5mm}
\noindent
is its rest-frame decay width. 
 
The uncertainty in the momentum of a particle produced in the decay is 
approximately given by the reciprocal of its coordinate uncertainty 
$\sigma_x$, which is essentially its velocity times the lifetime of the 
parent particle. Thus, for charged leptons produced in pion decay 
\be
\sigma_p\simeq [(p/E)\tau_\pi]^{-1} = (E/p)\Gamma_\pi\,.
\label{sigmap}
\ee

\vspace*{2.0mm}
\noindent
{}From eq.~(\ref{sigmae}) it then follows that the two terms in the square 
brackets in eq.~(\ref{sigmam1}) are approximately equal, and one finally gets

\be
\sigma_{m^2}\simeq 2\sqrt{2}\, E \sigma_E\,. 
\label{sigmam2}
\ee

\vspace*{2.0mm}
\noindent

It should be noted that $\sigma_{m^2}$, being the uncertainty of a 
Lorentz-invariant quantity, must itself be Lorentz invariant. 
Our estimate (\ref{sigmam2}) satisfies this condition. Indeed,  
when going from the pion's rest frame to the laboratory frame, the 
energies of the emitted leptons averaged over the directions of their 
rest-frame velocities with respect to the pion boost direction scale 
as $E\to \gamma E$. Together with eq.~(\ref{sigmae}), this proves Lorentz 
invariance of (\ref{sigmam2}).

The uncertainty in the charged-lepton mass determination in pion decay 
can therefore be estimated in pion's rest frame:
 
\be 
\sigma_{m^2}~\simeq~ 2 \sqrt{2} \,\bar{E}\, \Gamma_\pi^0 
~\simeq~2\sqrt{2}
\cdot 90~{\rm MeV}\cdot 2.5\cdot 10^{-8}~{\rm eV}~\simeq~6.4~\,{\rm eV}^2 
\,,
\label{estim1}
\ee

\vspace*{2.5mm}
\noindent
where $\bar{E}\simeq 90$ MeV is the average energy of charged leptons 
produced in pion decays at rest. This has to be compared with 
$m_\mu^2-m_e^2\simeq (106~{\rm MeV})^2$; obviously, the condition in 
eq.~(\ref{cond2}) is satisfied with a huge margin, which means that different 
mass-eigenstate charged leptons are emitted incoherently. Very similar 
estimates apply also to the decays of charged kaons.

Thus, we conclude that charged leptons born in the decays of pions or kaons 
(as well as in nuclear beta decays and in muon decays) are always emitted 
as mass eigenstates and not as coherent superpositions of different mass 
eigenstates because of their very large mass squared differences. Therefore 
even oscillations between the states $e_1$, ~$e_1$ and $e_3$ (or between 
different odor states $e_\beta$) discussed in the previous section are not 
possible -- these states just are not produced. Similar considerations apply 
to the absorption of charged leptons in weak-interaction processes
%%%%%%%%%%%%%%%%%%%%%%%%%%%%%
\footnote{If $\sigma_{pP}$ and $\sigma_{pD}$ are the momentum uncertainties 
associated with the production and detection processes, then 
eq. (\ref{sigmam1}) still holds with $\sigma_{p}$ being the effective 
uncertainty which is found from the relation $\sigma_p^{-2}=\sigma_{pP}^{-2}
+\sigma_{pD}^{-2}$, i.e. $\sigma_{p}$ is dominated by the smallest of the 
two uncertainties. The same applies to the energy uncertainties. These results 
can be obtained from the wave packet treatment of the production and detection 
processes and are similar to the corresponding results for neutrinos (see, 
e.g., \cite{Giunti,Beuthe}).}.
%%%%%%%%%%%%%%%%%%%%%%%%%%%%%

\vspace*{2.0mm}
Does this conclusion hold for all conceivable weak processes? The energies 
and momenta of charged leptons produced in pion and kaon decays are relatively 
small because of the small mass of the decaying particle, which implies small 
phase-space volumes available for the decay products. This (together with 
the chiral suppression which requires that the decay amplitudes be 
proportional to the lepton's mass) also explains relative smallness of the 
pion and kaon decay widths, which determine the uncertainties of the energies 
and momenta of the produced charged leptons. Altogether this results in rather 
small uncertainties $\sigma_{m^2}$ of the lepton masses and ensures that the 
condition (\ref{cond2}) is satisfied.   

Let us now consider $W$-boson decays $W^\pm \to l_a^\pm \nu$ ($l_a=e,\,\mu,\,
\tau$), which are characterized by large phase-space volumes and also do not 
suffer from the chiral suppression. The partial decay widths for such decays 
are  
\be
\Gamma_{W\to l_a\nu}^0\simeq \frac{G_F m_W^3}{6\sqrt{2}\pi}\simeq 
230~\,{\rm 
MeV}\,,
\label{gammaW}
\ee  
where $G_F$ is the Fermi constant and $m_W\simeq 80.4$ GeV is the $W$-boson 
mass. For $W$ decay at rest we therefore have the following 
estimate for the uncertainty of the charged lepton mass
\footnote{One might argue that the total $W$-boson width $\Gamma_W^0$ rather 
than the partial widths of $W^\pm \to l_a^\pm \nu$ decays should be used 
in this estimate. This would increase the estimate in eq.~(\ref{estim2}) 
by about a factor of 10, strengthening our conclusions. 
}:
\be 
\sigma_{m^2}~\sim~2\sqrt{2}\, E \sigma_E~\simeq~2\sqrt{2}\cdot 40~{\rm 
GeV}\cdot 230~{\rm MeV}~\simeq~(5~{\rm GeV})^2 \,.
\label{estim2}
\ee

\noindent
Thus, we have  
\be
\sigma_{m^2}
\gg m_\mu^2-m_e^2\,, \qquad
\sigma_{m^2}> m_\tau^2-m_\mu^2\simeq (1.77~{\rm GeV})^2\,,
\label{estim3}
\ee

\noindent
which means that all three charged leptons are produced coherently in $W^\pm$ 
decays. Since $\sigma_{m^2}$ is a Lorentz-invariant quantity, the same 
estimates (\ref{estim2}), (\ref{estim3}) and the same conclusion apply also 
for $W^\pm$ decays in flight. Thus, charged leptons are 
produced in $W^\pm\to l^\pm \nu$ decays as coherent superpositions of $e$, 
$\mu$ and $\tau$.

Does this mean that one can observe oscillations of such charged leptons 
if the detection process is also coherent? 
For the observability of the charged lepton oscillations it is not sufficient 
that the lepton state be produced as a coherent superposition of mass 
eigenstates; the emitted state should also preserve its coherence until 
it is detected. The coherence loss can occur for mixed states because of the 
finite effective spatial size $\sigma_x$ of the wave packet describing the 
propagating state. Since different mass-eigenstate components of the mixed 
state propagate with different group velocities $v_g=\partial E/\partial p$, 
after the coherence time 
\be
t_{\rm coh}\simeq \frac{\sigma_x}{\Delta v_g}
\label{tcoh}
\ee 

\noindent
the wave packets describing the individual mass eigenstates separate and 
can no longer interfere, which means that the state loses its coherence. 

Let us now estimate the coherence time for the charged lepton states produced 
in $W^\pm\to l^\pm \nu$ decays at rest and the corresponding coherence length  
$x_{\rm coh}$ (which for relativistic leptons coincides with the coherence 
time). The maximum coherence length corresponds to the minimum group 
velocity difference,
\be
(\Delta v_g)_{\min} =
\frac{p_e}{E_e}-\frac{p_\mu}{E_\mu}\simeq 
2\frac{m_\mu^2-m_e^2}{m_W^2}\,.
\label{Dvg}
\ee

\noindent
For the maximum coherence length we therefore find from eqs.~(\ref{tcoh}), 
(\ref{gammaW}) and (\ref{Dvg}) 

\be
(x_{\rm coh})_{\rm max}\simeq [\Gamma_{W\to l_a\nu}^0 (\Delta v_g)_{\min}]^{-1}
\simeq \frac{3\sqrt{2}\pi}{G_F m_W (m_\mu^2-m_e^2)}\simeq 2.5\times 
10^{-8}~{\rm cm}\,.
\ee

\vspace*{2.5mm}
\noindent
Thus, even though charged leptons are emitted in $W^\pm \to l^\pm \nu$ decays 
as coherent superpositions of mass eigenstates, they lose their coherence 
upon propagating only $\sim 10^{-8}$ cm from their birthplace, i.e. over 
interatomic distances. This means that coherent effects in the $l^\pm$ 
production are unobservable, and for all practical purposes one can consider 
the charged leptons produced in $W^\pm \to l^\pm \nu$ decays at rest to be 
an incoherent mixture of $e$, $\mu$ and $\tau$. 

What about $W^\pm \to l^\pm \nu$ decays in flight? Let $\gamma$ be the Lorentz 
factor of $W^\pm$. The minimum group velocity difference of the produced 
charged leptons $(\Delta v_g)_{\min}\simeq \Delta m_{\mu e}^2/2E^2 \equiv 
(m_\mu^2-m_e^2)/2E^2$ and the partial decay width of $W^\pm$ scale with 
$\gamma$ as  
\be
(\Delta v_g)_{\min} \to \gamma^{-2} (\Delta v_g)_{\min} 
\,,\qquad 
\Gamma_{W\to l_a\nu}^0\to \gamma^{-1} \Gamma_{W\to l_a\nu}^0 \,. 
\label{scale1}
\ee

\noindent
Therefore the maximum coherence length scales as 
\be
(x_{\rm coh})_{\rm max} \to \gamma^3 (x_{\rm coh})_{\rm max} \,.
\label{scale2}
\ee

\noindent
In order for $(x_{\rm coh})_{\rm max}$ to be, say, larger than 1 $m$, one 
would need $\gamma \gtrsim 1600$, or $E_W \gtrsim 130$ TeV, which is far 
above presently feasible energies. 

It is easy to see that the condition of having a coherent emission of charged 
leptons in a decay process and the condition that the leptons keep their 
coherence over a macroscopic distance $L$ tend to put conflicting constraints 
on the size $\sigma_x$ of the charged leptons' wave packet. Indeed, as 
follows from eqs. (\ref{cond3}) and (\ref{sigmam2}), the first condition 
requires 
\be
\sigma_x\sim \sigma_p^{-1}\simeq \sigma_E^{-1} < (\Delta m_{\mu e}^2/
2\sqrt{2} E)^{-1}\,,
\label{cond4}
\ee
whereas the second one yields
\be
\sigma_x > (\Delta v_g)_{\rm min} L\simeq  (\Delta m_{\mu e}^2/2E^2)\,L\,,
\label{cond5}
\ee
in accordance with eqs. (\ref{tcoh}) and (\ref{Dvg}). 
To reconcile the upper and lower limits on $\sigma_x$ given in eqs. 
(\ref{cond4}) and (\ref{cond5}), $L$ must satisfy 
\be
L < \frac{4\sqrt{2}\,E^3}{(\Delta m_{\mu e}^2)^2} ~\simeq ~8.9\times 
10^{-10}\,\left(\frac{E}{\rm GeV}\right)^3~{\rm cm}\,.
\label{cond6}
\ee
Note that this condition is independent of the size of the wave packet 
$\sigma_x$ and therefore of the decay width of the parent particle. From 
(\ref{cond6}) it follows that in order for charged leptons to be born  
coherently and keep their coherence over a distance $L\gtrsim 1\,m$ 
they should have energies greater than at least 4.8 TeV. A condition 
similar to that in eq.~(\ref{cond6}) exists also for neutrino oscillations, 
but in that case it is much easier to satisfy because of the smallness of 
neutrino mass squared differences. In particular, for a baseline $L\gtrsim 
1\,km$ one would only need neutrino energies $E_\nu\gtrsim 20$ eV. 

It should be stressed that the condition (\ref{cond6}) (and the similar 
condition for neutrinos) is necessary but in general not sufficient 
for a mixed state to be coherently produced and maintain its coherence 
over the distance $L$: it only ensures the consistency of the conditions 
(\ref{cond4}) and (\ref{cond5}) but not their separate fulfilment.  

%%%%%%%%%%
Several comments are in order. First, 
%%%%
%It should be noted that 
%%%%
a sufficiently coherent detection can improve the 
overall coherence of the total lepton production -- propagation --  detection 
process \cite{Kiers}. 
%%%%%%%%%%%%%%%%%%%%%%%%%%%%%%%%%%%%%%%%%%%%%%%
%If $\sigma_{pP}$ and $\sigma_{pD}$ are the momentum uncertainties 
%associated with the production and detection processes, then 
%eq. (\ref{sigmam1}) still holds with $\sigma_{p}$ being the effective 
%uncertainty which is found from the relation $\sigma_p^{-2}=\sigma_{pP}^{-2}
%+\sigma_{pD}^{-2}$, i.e. $\sigma_{p}$ is dominated by the smallest of the 
%two uncertainties. The same applies to the energy uncertainties. These results 
%can be obtained from the wave packet treatment of the production and detection 
%processes and are similar to the corresponding results for neutrinos (see, 
%e.g., \cite{Giunti,Beuthe}). The above means, in particular, that even if the 
%%%%%%%%%%%%%%%%%%%%%%%%%%%%%%%%%%%%%%%%%%%%%%%
In particular, even if the wave packets have already separated, they can still 
overlap and interfere in the detector if their separation is not too large and 
if the detection process is sufficiently coherent (i.e., lasts a sufficiently 
long time). In that case the separated wave packets arrive at the detector 
before the detection process is over. The coherence length is therefore the 
distance over which the wave packets separate to such an extent that they can 
no longer overlap in the detector. Our discussion of the separation of wave 
packets earlier in this section still holds if one understands by $\sigma_x$ 
the effective wave packet size, $\sigma_x=\sigma_p^{-1}\equiv (\sigma_{pP}^{-2}+
\sigma_{pD}^{-2})^{1/2}$, which takes the detection process into account  
(see footnote 5). 
In our numerical estimates we were assuming that the coherence of the 
detection process is not too different from that of the production process.  
%%%%%%%%%%

Second, in our discussion of the loss of coherence caused by the wave packet 
separation we have assumed that the size of the wave packet does not change 
with time, i.e. neglected the wave packet spreading. Such a spreading in  
general occurs when the group velocity depends on the particle's momentum 
(i.e. in the presence of dispersion). This is, in particular, the case 
for free relativistic massive particles, for which $\partial v_g/\partial 
p=m^2/E^3$. The asymptotic (large-$t$) spreading velocity is then 
$v_\infty = m^2/(E^3 \sigma_x)$ 
\footnote{This expression can be readily obtained from the general 
formulas given in Chapter 3 of \cite{Bohm}.
}.
The spreading increases the spatial size of the wave packets and 
therefore tends to counter the effect of the wave packet separation. 
The coherence can be recovered at large enough times provided that the 
asymptotic spreading velocity is larger than the difference of the group 
velocities:
\be
v_\infty = \frac{m^2}{E^3 \sigma_x} > \frac{|\Delta m_{ab}^2|}{2E^2}\,
\qquad (a, b = e,\,\mu,\,\tau)\,.
\label{asymp}
\ee
{}From $\sigma_x^{-1}\sim \sigma_p\simeq \Gamma$ and the fact that 
$|\Delta m_{ab}^2|\simeq {\rm max}(m_a^2,\,m_b^2)$ it follows that the 
condition (\ref{asymp}) reduces to the following inequality between the decay 
width of the parent particle and the energy of the produced charged lepton 
state:
\be
\Gamma \gtrsim E/2\,.
\label{cond}
\ee   
In reality this condition is never satisfied, which justifies our neglect 
of the wave packet spreading.

We have found that the charged lepton states born in the $W^\pm$ decays are 
produced coherently and can maintain their coherence up to macroscopic 
distances provided that $E_W\gtrsim 100$ TeV. However, as follows from the 
discussion in sec. 3, for these coherence effects 
to be experimentally observable  the following two conditions have to be 
satisfied: (i)~at the production, an accompanying neutrino must be detected, 
thus providing a measurement of the composition of the emitted charged 
lepton 
state.  Moreover, this neutrino must not be a flavor eigenstate $\nu_e$, 
$\nu_\mu$ or $\nu_\tau$ (otherwise the flavor of the produced charged lepton 
would be measured, so that it would be either $e$ or $\mu$ or $\tau$ but not 
their coherent superposition); (ii) the detection process should be able to 
discriminate between different coherent superpositions of charged leptons. 
Obviously, the standard charged-current weak interactions cannot meet these 
two conditions: the absorption of a neutrino state different from a flavor 
eigenstate would be accompanied by the emission of a mixed charged-lepton 
state, which would again have to be identified by its charged-current 
interaction, leading to the emission of the same mixed neutrino state. To 
break the circle, new interactions are necessary, and therefore we turn to 
possible new physics effects now.

\section{New physics?}

Assume very heavy sterile neutrinos $N_i$ exist (as required, e.g,, by the 
seesaw mechanism of neutrino mass generation) and consider their decay  
into a charged lepton and charged Higgs boson: 
\be
N_i \to e_i^- + \Phi^+ \,.
\label{dec1}  
\ee
This would also require the existence of an extra Higgs boson doublet  
because the charged component of the standard model Higgs is eaten up by the 
$W^\pm$ bosons through the Higgs mechanism 
\footnote{
An alternative possibility, to 
have the decay (\ref{dec1}) in the early universe above the electroweak 
symmetry breaking temperature, is of no interest to us: even though the 
charged component of the standard model Higgs would be physical in that case, 
the charged leptons would be massless and so would not oscillate.}. 
The decays in eq.~(\ref{dec1}) are caused by the Yukawa coupling Lagrangian 
\be
{\cal L}_{Y}=Y_{ai}\bar L_a N_{Ri} \Phi ~+ ~h.c.\,,
\label{Lag}
\ee
where $L_a=(\nu_{La},\, e_{L a})^T$ are the $SU(2)_L$ doublets of the left 
handed lepton fields. We work in the basis where the mass matrices of heavy 
sterile neutrinos and charged leptons have been diagonalized; the Yukawa 
coupling matrix $Y_{ai}$ is in general not diagonal in this basis, so that 
in the decay of a mass-eigenstate sterile neutrino $N_i$ any of the three 
charged leptons $e_a=e,\,\mu,\,\tau$ can be produced.  
We want to find out under what conditions the produced charged lepton state 
$e_i$ in eq.~(\ref{dec1}) is a coherent superposition of the mass eigenstates
$e_a$, in which case it is given by
\be
|e_i\rangle=[(Y^\dag Y)_{ii}]^{-1/2} \sum_a Y^\dag_{ia}\, |e_a\rangle\,, 
\label{e_i}
\ee

\vspace*{-1.5mm}
\noindent
and how long this state can maintain its coherence.

Neglecting the Higgs boson and charged lepton masses compared to the mass of 
the sterile neutrino $M_i$, for the rest-frame decay width of $N_i$ we find 
\be
\Gamma_i^0\simeq \alpha_i M_i\,,\qquad {\rm where} \qquad\,
\alpha_i \equiv \frac{(Y^\dag Y)_{ii}}{16\pi}\,.
\label{wid1}
\ee

\noindent
We now apply the arguments of the previous section to the decay (\ref{dec1}). 
The condition (\ref{cond3}) which has to be satisfied in order for the charged 
lepton state to be produced as a coherent superposition of $e$, $\mu$ and 
$\tau$ reads

\be
2\sqrt{2} \,E\, \Gamma_i^0\simeq 2\sqrt{2}\, (M_i/2)\,\alpha_i M_i > 
{\rm max}\{m_\mu^2-m_e^2, ~m_\tau^2-m_\mu^2\}\,,
\label{cond3a}
\ee
or
\be
\alpha_i > 2.2\, (M_i/{\rm GeV})^{-2} \,.
\label{cond3b}
\ee
{}From eq.~(\ref{tcoh}) we find the coherence length for the emitted 
charged lepton state:
\be
x_{\rm coh}\simeq \frac{M_i^2}{2 \Gamma_i^0 (m_\tau^2-m_\mu^2)}
\simeq 3.1\times 10^{-15}\;\alpha_i^{-1} \frac{M_i}{\rm GeV}\,~{\rm cm}\,.
\label{xcoh2}
\ee 
{}From eq.~(\ref{cond3b}) it then follows that 
\be
x_{\rm coh} < 1.4\times 10^{-15}\;{\rm cm} ~(M_i/{\rm GeV})^3\,.
\label{xcoh2a}
\ee 
Thus, for the charged lepton state to maintain its coherence over the distance 
of $\sim 1\;m$, the sterile neutrino must have the mass $M_i\gtrsim 400$ TeV. 
Eq.~(\ref{cond3b}) then implies that the Yukawa couplings $Y_{ij}$ must 
satisfy $(Y^\dag Y)_{ii}\gtrsim 1.3\times 10^{-11}$. If only $e$ and 
$\mu$ are to be produced coherently, a significantly milder lower limit 
on the sterile neutrino mass results: $M_i\gtrsim 10$ TeV, whereas for the 
Yukawa couplings one gets the constraint $(Y^\dag Y)_{ii}\gtrsim 8.5\times 
10^{-11}$. Note that for $N_i$ decay in flight the right hand side of 
eq.~(\ref{xcoh2a}) has to be multiplied by $\gamma^3$, which amounts 
to replacing the factor $(M_i/{\rm GeV})^3$ there by $(E_i/{\rm GeV})^3$. 
Thus, for $N_i$ decays in flight the condition of macroscopic coherence length 
puts a lower bound only on the energy of the sterile neutrinos, so that they 
can be relatively light. 

If the condition (\ref{cond3b}) for the coherent creation of the charged 
lepton state in the decay (\ref{dec1}) is satisfied and this state is 
detected through the inverse decay process before it loses its coherence, 
it may exhibit oscillations: a mass eigenstate sterile neutrino $N_j$  
different from $N_i$ can be produced in the detection process, meaning 
that the state $e_i$ has oscillated into $e_j$. The measurement of the 
``flavor'' of the originally produced $e_i$, i.e. of its composition with 
respect to the mass eigenstates $e$, $\mu$ and $\tau$ (as given in 
eq. (\ref{e_i})), is provided by the fact that the decaying sterile neutrino 
is a mass eigenstate.  

In our discussion of the decay (\ref{dec1}) we were assuming that the 
sterile neutrinos $N_i$ are heavier than the charged Higgs boson $\Phi$.  
If $\Phi$ is heavier than $N_i$, then decays 
\be
\Phi^\pm \to N_i + e_i^\pm  \
\label{dec2}  
\ee
are possible. This case can be analyzed quite analogously. Sterile neutrinos, 
being very heavy, are either emitted incoherently or lose their coherence 
almost immediately, providing a measurement of the ``flavor'' of the charged 
lepton state. Charged leptons then would be able to oscillate, leading to a 
non-zero probability of the emission or absorption of a different sterile 
neutrino mass eigenstate $N_j$ in the processes ~$e_j^\pm + \Phi^\mp \to N_j$ 
~or $e_j^\pm + N_j\to \Phi^\pm$. Thus, in the cases of decays (\ref{dec1}) 
and (\ref{dec2}) we have the roles of neutrinos and charged leptons reversed 
as compared to the usual situation because of sterile neutrinos being much 
heavier than the charged leptons. 

\section{Charged lepton oscillation lengths and averaging \\ over the 
source/detector size}

Up to now in our discussion we have been only considering the coherence 
properties of individual charged lepton states produced in decays of a single 
particle. However, in real experiments one normally has to deal with beams 
originating from the decays of the parent particles confined within a certain 
source volume, and the coordinate of the production point is usually only 
known with an uncertainty of the order of the (macroscopic) size of the source 
$L_S$. Likewise, the coordinate of the detection point is only known with the 
uncertainty of the order of the detector size $L_D$. In calculating the event 
rates one has to integrate over the coordinates of the production and detection 
points within their respective allowed spatial regions. Thus, the effective 
uncertainties of the coordinates of the production and detection points of 
charged leptons are usually much larger than the corresponding intrinsic 
quantum-mechanical uncertainties. 
As we shall see, because of this the requirement of macroscopic coherence 
lengths of charged leptons puts a very stringent lower bound on their 
energies. This bound stems from the condition of no averaging of the charged 
lepton oscillations over the lengths of the source and detector and is  
actually more stringent than the one coming from the condition of no wave 
packet separation.

If the charged lepton oscillation length $l_{\rm osc}$ is much smaller than 
the size of the source in the direction of the beam $L_S$, then the 
integration over the production point would average out the interference 
terms in the squared modulus of the amplitude of the process. The same is 
also true for the integration over the detection point provided that 
$l_{\rm osc} \ll L_D$. The absence of the interference terms would mean 
that the coherence effects in the charged lepton states are unobservable, 
and in each event a certain mass-eigenstate charged lepton is emitted or 
absorbed with its respective probability. 

Let us now estimate the energies $E_0$ of the decaying parent particle that 
are necessary for $l_{\rm osc}$ to take macroscopic values. The maximum 
oscillation length (corresponding to the smallest mass squared difference) is

\be
(l_{\rm osc})_{\rm max}\,=\,\frac{2\pi}{|E_\mu-E_e|}\simeq 2.5~m\; 
\frac{[(E_0/2)\,({\rm MeV})]}{\Delta m_{\mu e}^2\,({\rm eV}^2)}\,\simeq \,
1.1 \times 10^{-11} (E_0/{\rm GeV})~cm\,.
\label{losc}
\ee

\vspace*{1.5mm}
\noindent
Therefore in order to have, e.g., $l_{\rm osc}\gtrsim 1\,m$ one would need 

\be
E_0\gtrsim 9\times 10^{12}~{\rm GeV}\,,
\label{beyond}
\ee

\vspace*{1.5mm}
\noindent
which is far above the experimentally accessible energies (except, probably, 
for the highest-energy cosmic rays).

\section{Discussion and conclusions}

We have studied the coherence properties of the charged lepton states produced 
in weak-interaction processes and demonstrated that in those cases when the 
production of more than one type of mass-eigenstate charged leptons is 
kinematically allowed, the charged lepton states are either produced 
as incoherent mixtures of $e$, $\mu$ and $\tau$, or they lose their coherence 
over microscopic distances, except at extremely high energies, not accessible 
to present experiments. The reason for this difference between the coherence 
properties of neutrinos and charged leptons produced in weak-interaction 
processes is the enormous disparity between the masses of these two leptonic 
sectors of the standard model.  

We have also discussed charged lepton production in decays of heavy sterile 
neutrinos $N_i$ and demonstrated that in that case the oscillations between 
different coherent superpositions of $e$, $\mu$ and $\tau$ are possible, 
leading to potentially observable effects. The conditions for the 
observability of the oscillations of charged leptons produced in $N_i$ 
decays have been identified. 

We have studied three sources of decoherence of the charged lepton 
states: (i) lack of coherence at production; (ii) loss of coherence due 
to the wave packet separation, and (iii)~washout of coherence due to 
the averaging over the source and/or detector size. 
We have not considered the effects of $\mu$ and $\tau$ decays which could 
also cause loss of coherence of the charged lepton states. Except at extremely 
high energies, these decays occur on longer length  scales than the decoherence 
due to the wave packet separation. For example, in the case of $W^\pm$ decays,  
the decay length of the produced $\tau^\pm$ becomes shorter than the coherence 
length due to the wave packet separation $x_{\rm coh}$ only for $E_W\gtrsim 
2\times 10^6$ GeV, whereas the $\mu^\pm$ decay length becomes shorter than 
the corresponding $x_{\rm coh}$ only for $E_W\gtrsim 3\times 10^9$ GeV. 
For the charged leptons produced in the decays of heavy sterile neutrinos 
$N_i$ the instability of $\tau$ and $\mu$ becomes relevant for $E_i\gtrsim 
1.3\times 10^{6}$ GeV and $E_i\gtrsim 1.5\times 10^{10}$ GeV respectively.  

%%%%%%%%%%%%%%%%%%%%%%%%%
A necessary condition for the observability of the charged lepton oscillations 
is that they be emitted and detected as nontrivial linear superpositions 
of the mass eigenstates $e$, $\mu$ and $\tau$. For charged leptons produced in 
charged-current weak interactions the required ``measurement'' of the 
composition of their state can be achieved if, e.g, the accompanying neutrino 
is detected as a mass eigenstate. The measurements of the neutrino mass could 
in principle be performed through the time-of-flight techniques or through 
observation of decays of heavier neutrinos into lighter ones 
\footnote{The author is grateful to J. Rich for raising these points.}. 
However, it is easy to see that the current limits on neutrino masses and 
neutrino instability imply that the baselines (and flight times) necessary 
for such measurements are extremely large; this means that even if such 
measurements are performed, by the time they are done charged leptons will 
have already lost their coherence. 
%%%%%%%%%%%%%%%%%%%%%%%%%

The fact that charged leptons are always born in charged-current weak 
interactions as incoherent mass eigenstates or lose their coherence 
practically immediately has important consequences for neutrino oscillations. 
For neutrinos to oscillate, they should be produced and detected as 
well-defined coherent superpositions of mass eigenstates. This is trivially 
satisfied for neutrinos from $\beta$ decay, in which only electron-flavor 
neutrinos or antineutrinos are produced because the only charged leptons which 
can be emitted are $e^\pm$. The same is true for electron neutrinos or 
antineutrinos from $\mu\to e\nu\bar{\nu}$ decays, whereas the flavor of the 
other neutrino emitted in the same process is measured by the fact that the 
decaying particle (muon) is a mass eigenstate. However, in the decays such as 
$\pi^\pm\to l^\pm \nu$,  $K^\pm\to l^\pm \nu$ or $W^\pm\to l^\pm \nu$ the 
production of more than one charged lepton species is kinematically allowed. 
It is the lack of coherence of the produced charged lepton state or the loss 
of its coherence over microscopic distances that ensures that in each decay 
event a particular mass-eigenstate charged lepton is emitted and thus provides 
a measurement of the flavor of the associated neutrino. Only for this reason 
neutrinos emitted in such processes oscillate even when the associated charged 
lepton is not detected.

\vspace*{2mm}
Let us now briefly summarize our main conclusions:

\vspace*{-2mm}
\begin{itemize}

\item Charged leptons $e$, $\mu$ and $\tau$ do not oscillate into each other 
because they are mass eigenstates. Since in $\beta$ decays and muon decays 
the production of $\mu^\pm$ and $\tau^\pm$ is kinematically forbidden, the 
are no charged lepton oscillations associated with these processes.

\item Charged leptons born in $\pi^\pm$ and $K^\pm$ decays are produced  
incoherently, i.e. are either $\mu^\pm$ or $e^\pm$, but not their linear 
superpositions.  Therefore they do not oscillate. 

\item For charged leptons produced in $W^\pm$ decays the coherence production 
condition is satisfied. However, for $W^\pm$ decays at rest the coherence is 
lost over microscopic distances because of the wave packet separation. For 
decays in flight with $E_W\gtrsim 100$ TeV the coherence lengths can formally 
take macroscopic values; yet, the coherence effects in the charged lepton 
sector are unobservable even in this case because the standard charged-current 
weak interactions cannot provide a measurement of the composition of the 
initially produced as well as of the evolved charged lepton state.

\item
Charged lepton states produced in the decays of heavy sterile neutrinos 
can be coherent superpositions of $e$, $\mu$ and $\tau$. They can maintain 
their coherence over macroscopic distances provided that their energies 
exceed a few hundred TeV. Such charged lepton states could oscillate, and 
their oscillations could lead to observable consequences. 

\item Integration over the macroscopic sizes of the source and detector 
would wash out the effects of the oscillations of charged leptons unless the 
corresponding oscillation length exceeds the source and detector sizes in the 
direction of the beam, $L_{S}$ and $L_D$. The requirement of no washout for 
$L_{S}, L_D\gtrsim 1\,m$ puts a stringent lower bound on the energy of the 
decaying parent particle: $E_0\gtrsim 10^{13}$ GeV. 

\item Neutrinos produced in the processes in which the emission of more than 
one species of charged leptons is kinematically allowed oscillate even if the 
associated charged lepton is not detected, because the measurement of their 
flavor is provided by the decoherence of the associated charged lepton state.

\end{itemize}
Thus, the short answer to the question raised in the title of this paper is 
`no', at least if no new physics is involved. But even if the relevant new 
physics exists, an observation of the oscillations between different coherent 
superpositions of $e$, $\mu$ and $\tau$ would probably require extremely 
high energies, not accessible to current and most likely also to future 
experiments.

\vspace*{2mm}

The author is grateful to Joachim Kopp, Manfred Lindner and Alexei Smirnov 
for very useful discussions and to James Rich for an illuminating  
correspondence.

\end{document}